UDK 51-72

# Fitting of structural parameters to small angle neutron scattering data for nickel-chromium-aluminum alloy in frames of quantum mechanics and classic models of polydispersed spheres.

S.I.Potashev, V.P.Zavarzina, A.A.Afonin, R.A.Sadykov

*Institute for Nuclear Researches of Russian Academy of Sciences*

**Abstract**: Modified Yukawa's potential is used to fit model parameters of nucleus to the small angle thermal neutron scattering data on the nickel-chromium-aluminum alloy at the transferred momentum $Q$ and effective nucleus radius $R$ production satisfied by condition $QR \leq \hbar$. Analytical polydisperse sphere model is used to calculate a neutron scattering intensity and to determine most probable macroscopic sphere radius $R_0$ at $QR_0 \geq 3\hbar$. Combination of Gauss, Relay and Schulze-Zimm distributions can be used to fit model parameters to experimental data. Fast algorithm and program of fitting to data is proposed.

## 1. Introduction

To determine a substance structural parameters a small angle neutron scattering on a sample is investigated for neutron wavelength $\lambda = 4.5 \text{Å}$ which corresponds to neutron velocity 890 m/s. Two dimensional position-sensitive detector is used to detect a neutron scattered at small angles. As a result, a neutron scattering intensity dependence versus a transferred momentum $Q$ is measured. Here, $Q = mv \sin\theta$, where: $m$ — mass, $v$ - velocity, $\theta$ - scattering angle.

It is not possible to describe correctly experimental data of a thermal neutron scattering on nuclei for small values satisfied the condition $QR \leq \hbar$ in a frame of the classical approach [1]. Therefore, we use the model of nucleus with the modified Yukawa's potential in Born's approximation for a calculation. For large values satisfied the condition $QR_0 \geq 3\hbar$ a substance nanostructure can be explained in a frame of a classic model in which it is proposed that a sample consists of some crystal fractions of a sphere shape with the most probable radius $R_0$. Also, it is proposed that a sample contains some fractions of sphere crystals with the various radius distributions. Further, we calculate analytically a scattered neutron intensity $I$ versus a transferred momentum $Q$ with free parameters which are fitted to obtain the best accordance with the experimental data.

## 2. Quantum mechanics description of neutron scattering on nuclei

The scattering amplitude $f(Q)$ in Born's approximation

$$f(Q) = -\frac{m}{2\pi\hbar^2} \int V(r) \exp\left(-i\frac{\vec{Q}\vec{r}}{\hbar}\right) d^3x = \frac{2m}{\hbar^2} \frac{\hbar}{Q} \int_0^\infty V(r) \sin\left(\frac{Qr}{\hbar}\right) r\, dr , \qquad (1)$$

where: Q – transferred momentum. Introducing a variable $s = \frac{Q}{\hbar}$, expressed in the reciprocal length measurement units, we obtain

$$f(s) = -\frac{m}{2\pi\hbar^2} \int V(r) \exp(-i\vec{s}\vec{r}) d^3x = \frac{2m}{\hbar^2 s} \int_0^\infty V(r) \sin(sr) r\, dr . \qquad (2)$$

The nuclear potential of an each nucleus of the sample substance of interest is presented in Yukawa's form

$$V(r) = \frac{g}{r} \exp(-r/R_k) , \qquad (3)$$

where: index $k$ = 0, 1, 2 corresponds to *Ni, Cr, Al* nuclei. We introduce the effective nucleus radius $R_k = \xi A_k^{1/3} \, 10^{-13}$м, where $A_k$ - atomic weight of corresponded nucleus, $g$ - the effective nuclear charge and $\xi$ - a correcting multiplier. They are considered as free parameters which fitted to experimental data.

The differential cross section for an each nucleus

$$\frac{d\sigma_k}{d\Omega} = |f_k(s)|^2 = \frac{4g^2 \xi^4 m^2}{\hbar^4} \frac{P_k R_k^4}{\left(1 + \xi^2 R_k^2 s^2\right)^2}. \quad (4)$$

The measured scattering differential intensity

$$\frac{dI}{d\Omega} = I_0 L \left( n_0 \frac{d\sigma_0}{d\Omega} + n_1 \frac{d\sigma_1}{d\Omega} + n_2 \frac{d\sigma_2}{d\Omega} \right) = I_0 L \left( n_0 |f_0(s)|^2 + n_1 |f_1(s)|^2 + n_2 |f_2(s)|^2 \right) =$$
$$\frac{4 I_0 L \rho N_A m^2}{\hbar^4} \cdot g^2 \xi^4 \sum_{k=0}^{2} \frac{P_k}{A_k} \frac{R_k^4}{\left(1 + \xi^2 R_k^2 s^2\right)^2} \quad (5)$$

, where: $I_0$ - an incident intensity, $s$ — a transferred momentum, $m$ — a neutron mass, $N_A = 6,02 \cdot 10^{23}$ mol$^{-1}$, $\rho$ — a sample substance density, $n_k = \rho N_A \frac{P_k}{A_k}$ - a concentration of nucleus $k$, $P_k$ - a mass partition of nucleus $k$. The mass partitions in alloy *NiCrAl* are equals: $P_0$ = 0,58, $P_1$ = 0,39 and $P_2$ = 0,03 and atom weights are equals $A_0$ = 0,0587 kg/mol, $A_1$ = 0,052 kg/mol and $A_2$ = 0,027 kg/mol.

A curve of fitting to the begin part of a small angle neutron scattering experimental spectrum on the alloy *NiCrAl* by using a formula (3) and a Guinier approximation is shown in Fig.1. The most probable values of parameters are obtained: $g$ = 2.473 ± 0.05, $\xi$ = 28.3 ± 0.5. The most probable values of the Guinier sphere radius is equal $R$ = 67.2 ± 2,3A. However, in contrast to a classic approach the quantum mechanics gives the much better accordance to data of small angle scattering at $s \to 0$ and a scattered intensity intends to the finite magnitude at $s = 0$.

At small values $s$ we obtain the squared dependence versus $s$ as in the Guinier approximation which is valid for absolutely identical particles. This case is not realized for a distribution of various size macroscopic particles.

$$\frac{dI}{d\Omega} = \frac{4 I_0 L \rho N_A m^2}{\hbar^4} g^2 \xi^4 \sum_{k=0}^{2} \frac{P_k}{A_k} \frac{R_k^4}{\left[1 + 2(s\xi R_k)^2 + (s\xi R_k)^4\right]} \approx$$
$$\frac{4 I_0 L \rho N_A m^2}{\hbar^4} g^2 \xi^4 \left(\sum_{k=0}^{2} \frac{P_k R_k^4}{A_k}\right) \left(1 - 2s^2 \frac{\sum_{k=0}^{2} P_k R_k^6}{\sum_{k=0}^{2} P_k R_k^4}\right) \quad (6)$$

At large values $s$ we obtain the reciprocal four power dependence versus $s$ as in the Porod's low

$$\frac{dI}{d\Omega} = \frac{4 I_0 L \rho N_A m^2}{\hbar^4} g^2 \xi^4 \sum_{k=0}^{2} \frac{P_k}{A_k} \frac{R_k^4}{\left[1 + 2(s\xi R_k)^2 + (s\xi R_k)^4\right]} \approx \frac{4 I_0 L \rho N_A m^2}{\hbar^4} \frac{g^2}{s^4} \left(\sum_{k=0}^{2} \frac{P_k}{A_k}\right). \quad (7)$$

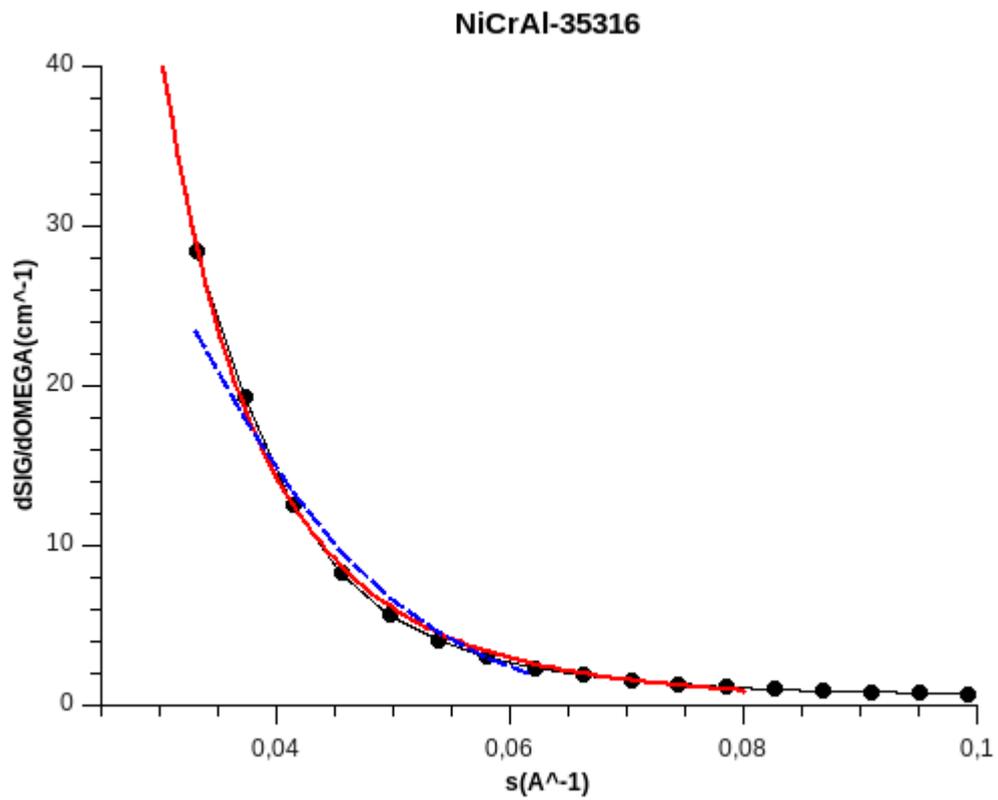

Fig.1. The experimental spectrum of a small angle scattering of neutron with wavelength 4,5A on a *NiCrAl* alloy and a fitting of begin part of the spectrum in Born's approximation with using the Yukava's potential (red solid curve) and in the Guinier low (dark blue curve).

## 3. Macroscopic model of spheres

The formula for scattering intensity is

$$I = \frac{d\Sigma}{d\Omega} = A \int_0^\infty f(R) \Phi^2(sR) \left(\frac{4\pi R^3}{3}\right)^2 dR, \quad A = n(\rho_u - \rho_M)^2, \quad (8)$$

where: $\Sigma$ - macroscopic cross section of a sample substance, $\Omega$ - a solid angle, $n$ - a molecule concentration, $\rho_u$ - a substance particle density, $\rho_M$ - a substance matrix density, $f(R)$ - a distribution function on a particle radius, $\Phi$ - a form factor, $s$ - a transferred momentum, $R$ - a particle radius.

Using formula from paper [1] for a sphere form factor and exchanging $t = sR$ we obtain the intensity distribution versus transferred momentum:

$$I(s) = \frac{16\pi^2}{s^6} \int_0^\infty f(R)[\sin(sR) - sR\cos(sR)]^2 dR = \frac{16\pi^2}{s^6} \int_0^\infty f(t)(\sin t - t\cos t)^2 dt. \quad (9)$$

### 3.1 Scattering intensity for the Schuze-Zimm distribution

The following normalized distribution density on a sphere radius is used:

$$df(R) = \frac{1}{\Gamma(k+1)\sigma^{k+1}} R^k e^{-\frac{R}{\sigma}} dR, \quad (10)$$

where the integral 860.07 in the book [2] is used.
The intensity of a scattering on a sphere for the Schulze-Zimm distribution

$$I(s) = \frac{1}{\Gamma(k+1)\sigma^{k+1}} \frac{(4\pi)^2}{s^6} \int_0^\infty R_k e^{-\frac{R}{\sigma}} [\sin(sR) - sR\cos(sR)]^2 dR. \quad (11)$$

We introduce dimensionless values: $a = 1/(s\sigma)$ and $t = sR$. In the conditions $k \in Z, k \geq 1$ we obtain $\Gamma(k+2) = (k+1)!$, $\Gamma(k+3) = (k+2)!$ and

$$I(s) = \frac{16\pi^2 a^{k+1}}{k! s^6} \int_0^\infty t^k e^{-at}(\sin t - t\cos t)^2 dt. \quad (12)$$

After non-complex but long transformations a solution of the integral is obtained [3]

$$\int_0^\infty t^k e^{-at}(\sin t - t\cos t)^2 dt = \frac{k!}{2a^{k+1}} - \frac{\Gamma(k+1)\cos[(k+1)\theta]}{2(a^2+4)^{\frac{k+1}{2}}} - \frac{\Gamma(k+2)\sin[(k+2)\theta]}{2(a^2+4)^{\frac{k+2}{2}}} + \frac{(k+2)!}{2a^{k+3}} + \frac{\Gamma(k+3)\cos[(k+3)\theta]}{2(a^2+4)^{\frac{k+3}{2}}}. \quad (13)$$

If to introduce values: $r = \sqrt{a^2+4}$, $\cos\theta = \frac{2}{r}$, $\sin\theta = \frac{a}{r}$, $\Theta_1 = (k+1)\theta$, $\Theta_2 = (k+2)\theta$, $\Theta_3 = (k+3)\theta$, the intensity can be written

$$I(s) = \frac{8\pi^2}{s^6} \left\{ 1 + \frac{k^2+3k+2}{a^2} - \left(\frac{a}{r}\right)^{k+1} \left[ \cos\Theta_1 + (k+1)\left(\frac{\sin\Theta_2}{r} - \frac{(k+2)\cos\Theta_3}{r^2}\right) \right] \right\}. \quad (14)$$

### 3.2 Scattering intensity for the Gauss distribution

The following normalized distribution density on a sphere radius is used:

$$df(R) = \frac{1}{\sqrt{2\pi}\sigma} e^{-\frac{(R-R_0)^2}{2\sigma^2}} dR, \quad \frac{1}{\sqrt{2\pi}\sigma} \int_0^\infty e^{-\frac{(R-R_0)^2}{2\sigma^2}} dR = 1. \quad (15)$$

The intensity for the Gauss distribution is calculated by the formula

$$I(s)=\frac{(4\pi)^2}{s^6}\int_0^\infty \frac{1}{\sqrt{2\pi}\sigma}e^{-\frac{(R-R_0)^2}{2\sigma^2}}[\sin(sR)-sR\cos(sR)]^2 dR. \quad (16)$$

A calculation is made in dimensionless variables: $t=sR$ and $t_0=sR_0$. We obtain

$$I(s)=\frac{(4\pi)^2}{s^7}\frac{\sqrt{s}}{\sqrt{2\pi}s\sigma}\int_0^\infty e^{-\frac{(t-t_0)^2}{2s^2\sigma^2}}(\sin t - t\cos t)^2 dt = \frac{16\pi^2}{\sqrt{2\pi}s^7}\frac{I_0}{\sigma}, \quad (17)$$

where: $s$ - transferred momentum, $R$ and $R_0$ - a current and the most probable sphere radius, $\sigma$ - a dispersion of a radios sphere distribution. Solving integral, we obtain

$$I(s)=\frac{16\pi^2}{\sqrt{2\pi}s^7}\cdot\frac{I_0}{\sigma}, \quad (18)$$

where

$$I_0 = \frac{s\sigma}{2}\sqrt{\frac{\pi}{2}}\left(1+t_0^2-2\sqrt{\frac{2}{\pi}}s\sigma t_0+s^2\sigma^2\right)+s^2\sigma^2[t_0\cos 2t_0-(1+s^2\sigma^2)\sin 2t_0]$$

$$-\frac{s\sigma}{2}\sqrt{\frac{\pi}{2}}e^{-2s^2\sigma^2}\left\{[1-t_0^2+s^2\sigma^2(3+4s^2\sigma^2)]\cos 2t_0+(1+2s^2\sigma^2)2t_0\sin 2t_0\right\}, \quad (19)$$

$$+\frac{s\sigma}{\sqrt{2}}F(s\sigma\sqrt{2})\left\{[1-t_0^2+s^2\sigma^2(3+4s^2\sigma^2)]\sin 2t_0-(1+2s^2\sigma^2)2t_0\cos 2t_0\right\}$$

where the function $F(y)$ is presented as the integral or series

$$F(y)=\frac{1}{2}\int_0^\infty e^{-\frac{t^2}{4}}\sin(yt)dt = \sum_{k=0}^\infty \frac{(-1)^k 2^k}{(2k+1)!!}y^{2k+1}. \quad (20)$$

### 3.3 Scattering intensity for the Relay distribution

The intensity versus the transferred momentum for a scattering on spheres with the Relay's distribution on radii

$$I(s)=\frac{16\pi^2}{s^6}\int_0^\infty \frac{R}{\sigma^2}e^{-\frac{R^2}{2\sigma^2}}[\sin(sR)-sR\cos(sR)]^2 dR, \quad \int_0^\infty \frac{R}{\sigma^2}e^{-\frac{R^2}{2\sigma^2}}dR=1. \quad (21)$$

After an exchange $t=sR$

$$I(s)=\frac{16\pi^2}{s^6}\int_0^\infty \frac{t}{s^2\sigma^2}e^{-\frac{t^2}{2s^2\sigma^2}}(\sin^2 t - 2t\sin t\cos t + t^2\cos^2 t)dt = \frac{16\pi^2}{s^6}I_0, \quad (22)$$

where

$$I_0=\frac{3}{4}+2s^2\sigma^2-s\sigma\left[\frac{5\sqrt{2}}{2}+s\sigma(3+4s^2\sigma^2)\right]F(s\sigma\sqrt{2}). \quad (23)$$

The intensity versus a transferred momentum is

$$I(s)=\frac{16\pi^2}{s^6}I_0=\frac{16\pi^2}{s^6}\left\{\frac{3}{4}+2s^2\sigma^2-s\sigma\left[\frac{5}{\sqrt{2}}+s\sigma(3+4s^2\sigma^2)\right]F(s\sigma\sqrt{2})\right\}. \quad (24)$$

## 4. A searching of local minimum of function $\chi^2$

We propose that scattering sphere particles consist of three fractions and their sizes following three types of distributions: Schulze-Zimm, Gauss or Relay. The function $\chi^2$ is written as

$$\chi^2 = \sum_{i=0}^{n-1} (I_i - D_i)^2 W_i = \sum_{i=0}^{n-1} \left( \sum_{j=0}^{2} N_j I_{ji} - D_i \right)^2 W_i, \quad I_i = \sum_{j=0}^{2} N_j I_{ji}, \qquad (25)$$

where

$N_j$ - the portion of particles with given distribution;
$I_{ji}$ - the calculated partial intensity for transferred momentum $S_i$;
$I_i$ - the calculated total intensity for transferred momentum $S_i$;
$D_i$ - the experimental total intensity for transferred momentum $S_i$;
$W_i$ - the weight coefficient for transferred momentum $S_i$;
$i$ - an index of transferred momentum;
$n$ - a number of measured experimental points;
$j$ - an index of particle distribution.

Unknown parameters are $N_j$ and

$R_j$ - the most probable particle radius for $j$-th distribution;
$\sigma_j$ - a particle dispersion for $j$-th distribution.

The derivatives of $\chi^2$ on these parameters are

$$\frac{\partial \chi^2}{\partial R_k} = 2 \sum_{i=0}^{n-1} \left( \sum_{j=0}^{2} N_j I_{ji} - D_i \right) W_i \sum_{j=0}^{2} N_j \frac{\partial I_{ji}}{\partial R_k}, \quad I_i = \sum_{j=0}^{2} N_j I_{ji}. \qquad (26)$$

The function $\chi^2$ is considered as a multidimensional surface in a space of $N_j$, $R_j$, $\sigma_j$. To determine the local minimum of $\chi^2$ the iteration on these parameters is made with some step.

Let us $m$ - the iteration number and magnitudes o $\chi^2$ f and its derivatives are known at previous iteration $m$. Let us $\chi^2$ on an each iteration is multiplied on some value $1-\beta$ where $\beta \in [0;1]$.
It follows the steps of following iteration are obtained

$$\Delta R_{k\,m} = \frac{\beta \chi_m^2}{\left( \frac{\partial \chi_m^2}{\partial R_k} \right)} \quad \Delta \sigma_{k\,m} = \frac{\beta \chi_m^2}{\left( \frac{\partial \chi_m^2}{\partial \sigma_k} \right)} \quad \Delta N_{k\,m} = \frac{\beta \chi_m^2}{\left( \frac{\partial \chi_m^2}{\partial N_k} \right)}. \qquad (27)$$

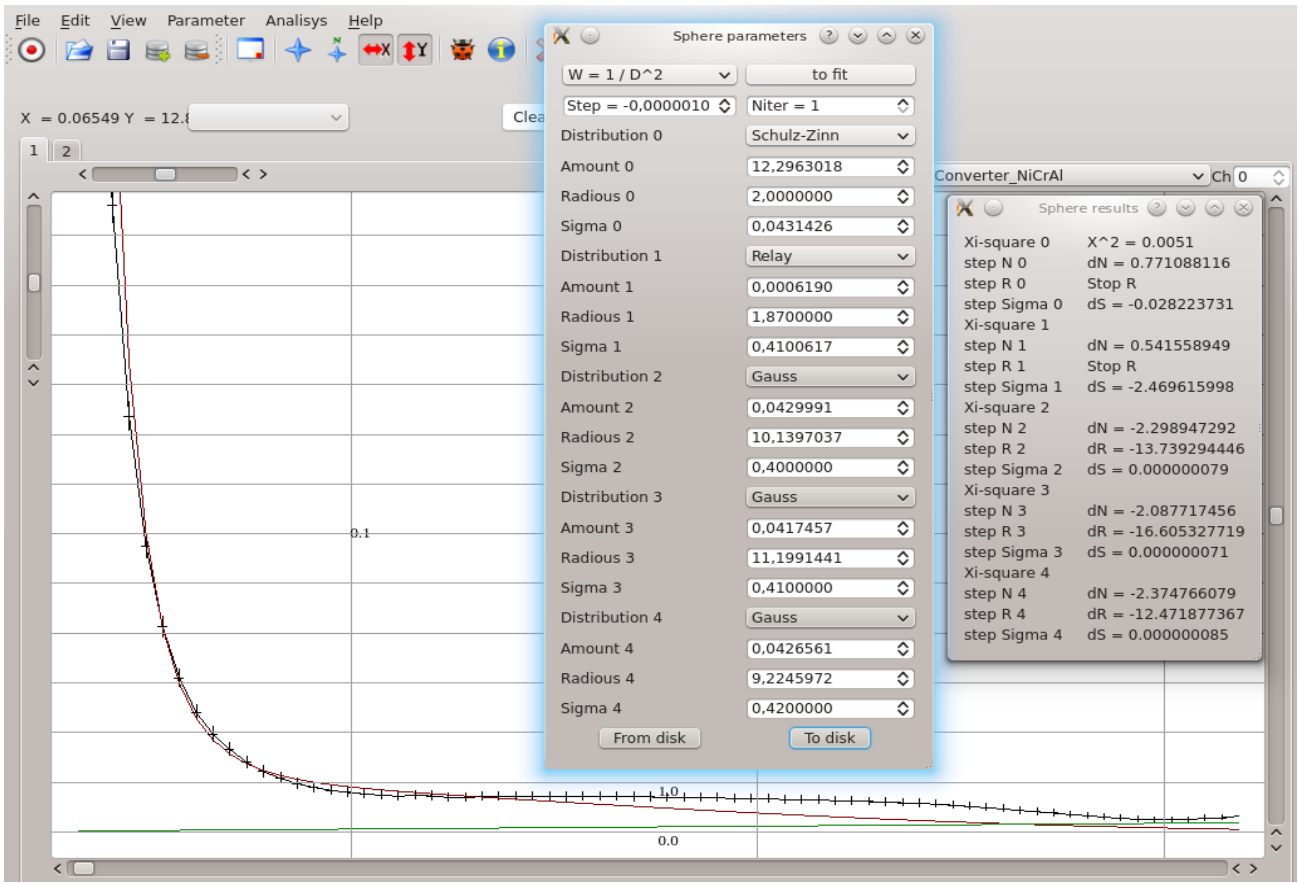

Fig.2. The experimental spectrum of small angle neutron scattering on the *NiCrAl* alloy sample at 4.5Å and the a fitting curve with using a combination of five fraction of spheres.

The $R_k$ step is $\Delta R_k = 0$ for the Schulze-Zimm and Relay distribution and therefore a fitting is not made on $R_k$. The weights is equal $W_i = \sqrt{D_i}$ for the Gauss distribution and they depends only experimental data.

## 6. Discussion

The experimental spectrum of small angle neutron scattering on the *NiCrAl* alloy sample at 4.5Å with the a fitting curve is shown in Fig.2. As a result, it was found that the main contribution in the experimental spectrum is due to the Schulze-Zimm fraction with the most probable sphere radius of 46Å. The distribution is shown in Fig.3. The second fraction is described by three Gauss distributions with the most probable sphere radius of 9, 10 and 11Å. Their combined distribution is shown in Fig.4. The third small fraction consists of particles of the Relay distribution with the most probable sphere radius of 5Å. The corresponded distribution is shown in Fig.5. The results of fitting are presented in table 1.

Table 1. The fitting results to experimental data.

| Distribution type | Contribution, cm$^{-1}$ | Radius, Å | Dispersion, Å | Power | Amplitude decay, Å$^{-1}$ |
|---|---|---|---|---|---|
| Schulze-Zimm | 12,2963018 | 46,35 | - | 2 | 0,0431426 |
| Relay | 0,0006190 | 5 | 0,4100617 | - | - |
| Gauss | 0,0429991 | 10,1397037 | 0.4 | - | - |
| Gauss | 0,0417457 | 11,1991441 | 0.41 | - | - |
| Gauss | 0,0426561 | 9,2245972 | 0.42 | - | - |

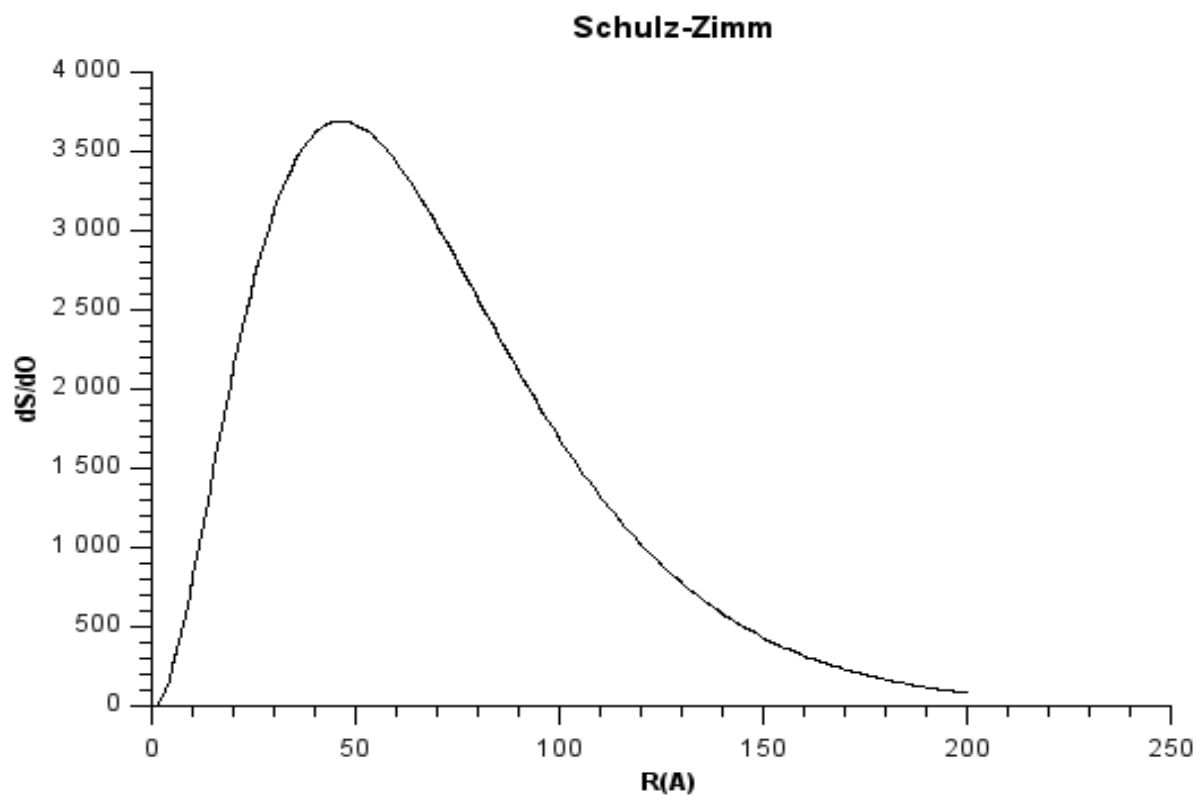

Fig.3. The sphere distribution on the radius obtained in the experimental data analysis. The most probable radius is equal 46Å. The power *k* is equal 2 in the Schulze-Zimm function.

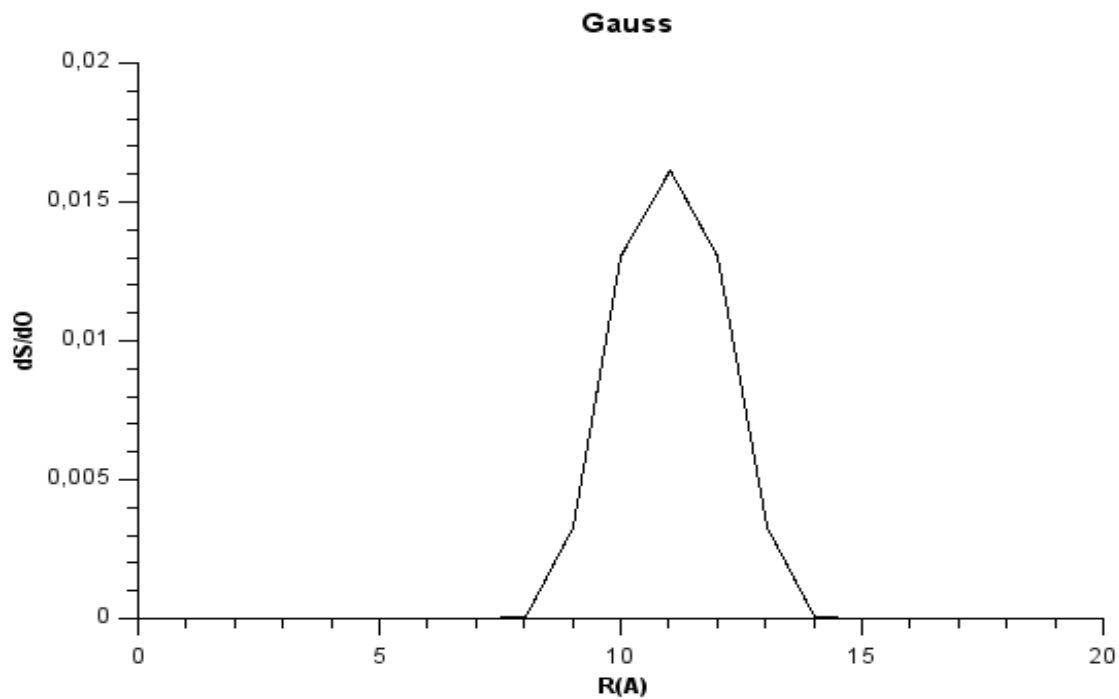

Fig.4. A combination contribution of three fractions of spheres of the Gauss distributions with the most probable radii 9Å, 10Å and 11Å.

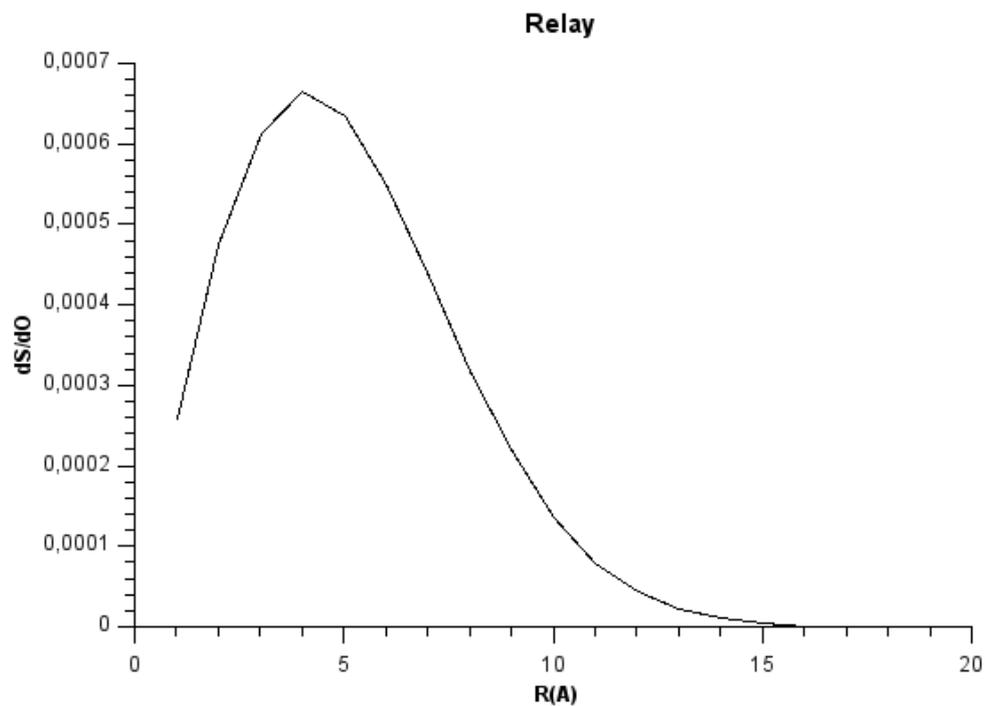

Fig.5. A contribution of the fraction of the Relay distribution with the most probable radius 5Å.

## 7. Acknowledgement

## 8. Conclusions

1. The quantum mechanics approach in the nucleus model with the modified Yukawa's potential for a calculation of neutron scattering intensity at $QR \leq \hbar$ is used. It describes much better the small angle neutron scattering spectrum at a small transferred momentum than the classic Guinier model;
2. The scattering intensity and its derivatives versus transferred momentum are analytically calculated in the sphere model for Schulze-Zimm, Gauss and Relay distributions at $QR_0 \geq 3\hbar$;
3. The fast algorithm and program for fitting is used for experimental data analysis;
4. An analysis of small angle neutron scattering data on the *NiCrAl* alloy is made and distributions on a radius are obtained by using this program.